\documentstyle[12pt]{article}

\newcommand{\sect}[1]{\setcounter{equation}{0}\section{#1}}

\newcommand{\EQ}{\begin{equation}}
\newcommand{\EN}{\end{equation}}
\newcommand{\bea}{\begin{eqnarray}}
\newcommand{\ena}{\end{eqnarray}}

\renewcommand{\a}{\alpha}
\renewcommand{\b}{\beta}
\renewcommand{\c}{\gamma}
\renewcommand{\d}{\delta}
\newcommand{\e}{\epsilon}
\newcommand{\la}{\lambda}

\newcommand{\shalf}{\frac{1}{2}}
\newcommand{\pa}{\partial}
\begin{document}

\topmargin 0pt
\oddsidemargin 5mm

\renewcommand{\Im}{{\rm Im}\,}
\newcommand{\NP}[1]{Nucl.\ Phys.\ {\bf #1}}
\newcommand{\PL}[1]{Phys.\ Lett.\ {\bf #1}}
\newcommand{\NC}[1]{Nuovo Cimento {\bf #1}}
\newcommand{\CMP}[1]{Comm.\ Math.\ Phys.\ {\bf #1}}
\newcommand{\PR}[1]{Phys.\ Rev.\ {\bf #1}}
\newcommand{\PRL}[1]{Phys.\ Rev.\ Lett.\ {\bf #1}}
\newcommand{\PTPS}[1]{Suppl.\ Prog.\ Theor.\ Phys.\ {\bf #1}}
\newcommand{\MPL}[1]{Mod.\ Phys.\ Lett.\ {\bf #1}}
\newcommand{\IJMP}[1]{Int.\ Jour.\ Mod.\ Phys.\ {\bf #1}}
\newcommand{\JP}[1]{Jour.\ Phys.\ {\bf #1}}
\renewcommand{\thefootnote}{\fnsymbol{footnote}}
\begin{titlepage}
\setcounter{page}{0}
\rightline{FERMILAB-PUB-91/228-T}
\rightline{OS-GE 20-91}
\rightline{Brown-HET-834}

\vspace{.5cm}
\begin{center}
{\Large BRST COHOMOLOGY AND PHYSICAL STATES IN 2D SUPERGRAVITY
COUPLED TO ${\hat c} \leq 1$ MATTER}
\vspace{1cm}

{\large Katsumi Itoh$^{1,2}$ and Nobuyoshi Ohta$^{1,3}$} \\
\vspace{1cm}
$^1${\em Fermi National Accelerator Laboratory \\
    P. O. Box 500, Batavia, IL 60510, U. S. A.}\\
\vspace{.5cm}
$^2${\em Department of Physics, Brown University \\
    Providence, RI 02912, U. S. A.}\\
\vspace{.5cm}
$^3${\em Institute of Physics, College of General Education,
Osaka University  \\  Toyonaka, Osaka 560, Japan} \\
\end{center}
\vspace{5mm}
\centerline{{\bf{Abstract}}}

We study the BRST cohomology for two-dimensional supergravity coupled
to $\hat c \leq 1$ superconformal matter in the conformal gauge. The
super-Liouville and superconformal matters are represented by free
scalar fields $\phi^L$ and $\phi^M$ and fermions $\psi^L$ and $\psi^M$,
respectively, with suitable background charges, and these are coupled
in such a way that the BRST charge is nilpotent. The physical states of
the full theory are determined for NS and R sectors. It is shown that
there are extra states with ghost number $N_{FP}=0,\pm 1$ for discrete
momenta other than the degree of freedom corresponding to the ``center
of mass", and that these are closely related to the ``null states" in
the minimal models with $\hat c<1$.

\vspace{2cm}
\end{titlepage}
\newpage
\renewcommand{\thefootnote}{\arabic{footnote}}
\setcounter{footnote}{0}

\sect{Introduction}

There has recently been significant progress in attempts to find
nonperturbative treatment of two-dimensional gravity and string theory.
In the matrix models, conformal field theories with central charges $c
\leq 1$ are successfully coupled to quantum gravity~\cite{BDG,BRE,PGB}
and partition functions as well as correlation functions have been
computed~\cite{GRO}. In particular, it has been found that there are
infinite number of extra states at discrete values of momenta other than
the degree of freedom corresponding to the ``center of mass" or
``tachyon'', but the nature and the role of these states in the theory
have not been fully understood yet.

To get more insight into the theory, it is clearly necessary to
understand these results from the viewpoint of the usual continuum
approach to the two-dimensional gravity: the Liouville theory~\cite{DDK}.
Most of the results in the matrix models have been confirmed in the
recent study of the Liouville theory~\cite{DDK,SPO}. In particular,
several groups have computed correlation functions and partition
functions to find again discrete states~\cite{GTW,BKL,GFK,GKL,POL}.
Some attempts to clarify the properties of these states have been made
in \cite{DGR,IT2,MMS,IMM}.

A first step toward full understanding of these states has recently
been taken in the BRST approach~\cite{LIZ,BMP,KOG,IKU}. In this approach,
physical states are characterized as nontrivial cohomology classes of
the BRST charge. Indeed, recent complete analysis of the BRST cohomology
has revealed that there are nontrivial physical states with ghost
numbers $N_{FP}=0, \pm 1$ at special values of momenta, corresponding
to the extra states found in the matrix models~\cite{LIZ,BMP}.

On the other hand, little is known for the supersymmetric case except
for Marinari and Parisi's proposal for supersymmetric matrix
models~\cite{MAP}. The two-dimensional supergravity coupled to ${\hat c}
 (\equiv \frac{2}{3}c)\leq 1$ superconformal matter reduces to coupled
super-Liouville theory and it correctly reproduces the scaling
dimensions~\cite{DHK}. However, no correlation functions have
been computed and the physical spectrum has not been clarified. In view
of the potential significance of superstring theory, it is very important
to understand the physical spectrum in two-dimensional supergravity
coupled to ${\hat c} \leq 1$ superconformal matter.

The purpose of this paper is to compute the BRST cohomology and identify
the physical spectrum for such a system. Our computation of the BRST
cohomology is quite analogous to the bosonic case~\cite{LIZ,BMP}. One
first decomposes the BRST charge~\cite{OHT,MIT,ITO,TKU} with respect to the
ghost zero modes. These are further decomposed according to a grading
of the Fock space, and we sort out the nontrivial cohomology classes
of the operator with lowest degree. In the critical superstring
\cite{OHT,MIT} as well as in the bosonic Liouville theory~\cite{LIZ,BMP},
there is a one-to-one correspondence between this nontrivial cohomology
and the cohomology of the total BRST charge. It turns out that this is
not true in general in the super-Liouville theory. We have carefully
examined which of these nontrivial states can be promoted to the
nontrivial cohomology classes of the full BRST charge for the
Neveu-Schwarz (NS) and Ramond (R) sectors. In this way, we show that
there are indeed nontrivial physical states for discrete values of
momenta, corresponding precisely to the ``null states" in the minimal
superconformal models~\cite{BKT,KMA}. The same results are also obtained
by using the decoupling mechanism based on the quartet structure with
respect to the BRST charge.

In sect. 2, we start by reviewing briefly the super-Liouville theory
coupled to ${\hat c} \leq 1$ matter. We use the free field realization
with a background charge for the matter sector~\cite{KMA,FF,FFR}.
We then study the BRST cohomology for the NS sector in sect. 3 and for
R sector in sect. 4, and identify the extra physical discrete states.
Sect. 5 is devoted to discussions. In particular, we point out the
close relationship of these extra states to the ``null states" in the
${\hat c}<1$ minimal models.

\sect{Super-Liouville theory coupled to ${\hat c} \leq 1$ matter}

In this section, we briefly summarize the super-Liouville theory
coupled to the $\hat c\leq 1$ matter theory for completeness.
This also serves to establish our notations and conventions.

In the conformal gauge, the matter and super-Liouville theories can
be realized by free superfields $\Phi^M$ and $\Phi^L$ which contains
scalar $\phi$ and fermionic fields $\psi$
\EQ
\Phi (\theta, z) = \phi (z) -i\theta\psi(z)
\EN
with the two-point functions
\EQ
<\phi (z) \phi (w)>=-\ln (z-w)\;\;\;, \hspace{5mm}
<\psi (z) \psi(w)>=\frac{1}{z-w}.
\EN
For the moment we will suppress the superscripts and describe a free
superfield realization, which is applicable to both the matter and
gravity sectors.

The super-stress tensor is given by
\bea
T(\theta,z) & = & -\frac{1}{2}D\Phi D^2\Phi - i \la
D^3\Phi\nonumber\\
& \equiv & \frac{1}{2} T_F+\theta T_B
\ena
where $D\equiv \partial_\theta + \theta \partial$ is the covariant
derivative. In terms of the component fields defined in eq.~(2.1), the
stress-energy tensor $T_B$ and super-current $T_F$ are expressed as
\bea
T_B & = & -\frac{1}{2} (\partial \phi )^2 -\frac{1}{2} \psi \partial
\psi - i \la \partial^2\phi , \nonumber\\
T_F & = & i \psi \partial \phi - 2\la \partial \psi
\ena
which satisfy the $N=1$ superconformal operator product with the
central charge $c=1 + \frac{1}{2} -12\la ^2$ or $\hat c =1 -8\la^2.$

The mode expansions are defined by
$$
\phi(z) = q - i(p-\la) \ln z + i \sum_{n\neq 0}\frac{\a_{n}}{n}
z^{-n}, \eqno(2.5a)
$$
$$
\psi (z) = \sum_n \psi_nz^{-n-\frac{1}{2}}\eqno(2.5b)
$$
with the commutation relations
\setcounter{equation}{5}
\bea
[ \a _n, \a_m] & = & n \d_{n+m,0},\;\;
[q,p]=i\;\;,\nonumber \\
\{\psi _n,\psi_m\} & = & \d_{n+m, 0}\;\;.
\ena
In eq.~(2.5b), the sum over $n$ is to be taken over half-odd-integers
for the NS sector and integers for the R sector.

The super-Virasoro generators are defined to be the Laurent
coefficients of the super-stress tensor. In terms of the mode
operators, they are given by
\bea
L_n & = & \frac{1}{2} \sum_m:\a_m\a_{n-m}:
+\frac{1}{4}\sum_m(2m-n):\psi_{n-m}\psi_m: +(n+1)\la\a_n, \nonumber\\
G_n & = & \sum_m\psi_{n-m}\a_m+(2n+1)\la \psi_n
\ena
where $\a_0\equiv p-\la$.
Note that the subscript $n$ to $G$ is half-odd-integer or integer
depending whether it is for the $NS$ or $R$ sector.

In the present case of super-Liouville theory with the background
charge $\la^L$ coupled to the matter with $\la^M$, we have two sets of
the above system. The total BRST charge is then
\bea
Q_B & = & \sum_n c_{-n}\left(L_n^M + L_n^L\right)-\frac{1}{2}
 \sum_n \c_{-n}\left(G_n^M+G_n^L\right)-\frac{1}{2}\sum_{n,m}
(n-m):c_{-n}c_{-m}b_{n+m}:\nonumber\\
& & + \sum_{n,m}\left(\frac{3}{2}
n+m\right):c_{-n}\c_{n+m}\b_{-m}: -\frac{1}{4}
\sum_{n,m}\c_n\c_mb_{-n-m}
\ena
where the sum is to be taken such that the subscripts to $G,\c$ and
$\b$ are half-odd-integers (integers) for NS (R) case and others are
integers. The commutation relations for the ghosts are
\EQ
\{c_n,b_m\} = [\c_n,\b_m] = \d_{n+m,0}.
\EN

The central charges for the matter and super-Liouville systems are
given by ${\hat c}^M = 1-8(\la^M)^2$ and ${\hat c}^L = 1 -8(\la^L)^2$,
respectively. Requiring that the total central charge add up to zero
or the BRST charge be nilpotent give ${\hat c}^M+{\hat c}^L-10=0 $ or
\EQ
(\la^M)^2 + (\la^L)^2 = -1.
\EN
Note that the conditions $\hat c ^M \leq 1$ and (2.10) mean that
$\la^M$ is real whereas $\la^L$ is pure imaginary.

The BRST charge can be decomposed with respect to the ghost zero
modes. For the NS sector
\EQ
Q_B = c_0L_0-b_0M_{NS} + d_{NS}
\EN
where\footnote{In what follows, it is understood that $n,m,\cdots$
take integer values whereas $r,s,\cdots$ take half-odd-integers
unless otherwise specified.}
\bea
L_0 & = & L_0^M + L_0^L+L_0^G \;, \nonumber\\
M_{NS} & = & \sum _{n\neq 0} n c_{-n}c_n  + \frac{1}{4} \sum
_r\c_{-r}\c_r, \nonumber\\
d_{NS} & = & \sum_{n\neq 0} c_{-n}
\left(L_n^M+L_n^L\right)-\frac{1}{2}
\sum_{nm(n+m)\neq 0} (m-n):c_{-m}c_{-n}b_{m+n}:\\
& & - \frac{1}{2} \sum_r \c_{-r}(G_r^M+G_r^L)
+\sum_{\stackrel{n\neq 0}{m}}
\left(\frac{3}{2} n+r\right):c_{-n}\c_{n+r}\b_{-r}:
-\frac{1}{4} \sum _{r+s\neq 0}\c_r\c_sb_{-r-s}.\nonumber
\ena
The nilpotency of $Q_B$ is equivalent to the following set of
identities:
\EQ
d_{NS}^2 = M_{NS}L_0, \hspace{5mm}
[d_{NS},L_0] = [d_{NS}, M_{NS} ] = [L_0, M_{NS}] =0\;\;.
\EN

For our purpose, it is convenient to define a set of generalized
momenta
\EQ
P^\pm (n) = \frac{1}{\sqrt 2}[(p^M +n \la^M)\pm i
(p^L+n\la^L)].
\EN
In particular, these give the ``lightcone-like" momenta for $n=0$
\EQ
p^\pm \equiv P^\pm (0)= \frac{1}{\sqrt 2} (p^M \pm ip^L).
\EN
We also define other lightcone-like variables by
\bea
q^\pm & = & \frac{1}{\sqrt 2}(q^M\pm i q^L),\;\;\; \a^\pm_n =
\frac{1}{\sqrt 2}(\a^M_n \pm i \a^L_n),\;\;\; \nonumber\\
\psi ^\pm_r & = & \frac{1}{\sqrt {2}} (\psi^M_r\pm i
\psi^L_r)\;\;\;,
\ena
which satisfy the commutation relations
\EQ
[q^\pm, p^\mp ] = i\;\;\;,\;\; [\a^\pm_m, \a^\mp_n] = m \d
_{n+m,0}\;\; ,\;\; \{\psi^\pm_r, \psi^\mp_s\} = \d_{r+s,0}.
\EN
Using these variables, the operators in eq.~(2.12) are cast into the
form
$$
L_0 = p^+p^- + \sum_{n\neq 0}:\a^+_{-n} \a^-_n:+\sum_r
r: \psi^+_{-r}\psi^-_r: +\sum_{n\neq 0} n: c_{-n}b_n: +\sum_r r:\b
_{-r}\c_r:
$$
$$
= p^+p^-+{\hat N}\eqno(2.18a)
$$
$$
d_{NS} = \sum_{n\neq 0} c_{-n}[P^+(n)\a^-_{n}+P^-(n)\a^+_n] +
\sum_{\stackrel{n,m\neq 0}{n+m\neq 0}}
:c_{-n}[\a^+_{-m}\a^-_{m+n}+
\frac{1}{2}(m-n)c_{-m}b_{m+n}]:
$$
$$
-\frac{1}{2} \sum_{n,r}(2r+n):c_{-n}\psi^+_{n+r}\psi^-_{-r}:
-\frac{1}{2}\sum_r\c_{-r}[P^+(2r)\psi ^-_r+P^-(2r)\psi^+_r] $$
$$
 - \frac{1}{2} \sum_{\stackrel{n\neq 0}{r}}
\c_{-r}(\psi^+_{r-n}\a^-_n+\psi^-_{r-n}\a^+_n)
+\sum_{\stackrel{n\neq 0}{r}}(\frac {3}{2}
n+r):c_{-n}\c_{n+r}\b_{-r}:
- \frac{1}{4} \sum_{r+s\neq 0}\c_r\c_s b_{-r-s}
\eqno(2.18b)
$$
\setcounter{equation}{18}
In writing down eq.~(2.18a), we have subtracted the intercept
$\frac{1}{2}$ which appears in rewriting the zero mode part of $L_0$
in terms of $p^+$ and $p^-$ using (2.10). This is necessary
in order to make the BRST charge (2.8) nilpotent~\cite{OHT,MIT}.

Similarly the BRST charge for the R sector is decomposed as
\EQ
Q_B= c_0L_0-b_0M_R-\frac{1}{2} \c _0 F + 2\b_0K+d_R-\frac{1}{4}
b_0\c^2_0
\EN
where $L_0$ is the same as (2.18a) with all the sum over integers and
\bea
M_R & = & \sum_{n\neq 0} (n c_{-n}c_n + \frac{1}{4}
\c_{-n}\c_n)\nonumber\\
F & = & p^+\psi^-_0 + p^-\psi ^+_0 + \sum_{n\neq 0}
(\psi^+_{-n}\a^-_n+\psi^-_{-n}\a ^+_n -n
c_{-n}\b_n+\c_nb_{-n})\nonumber\\
K & = & \frac{3}{4} \sum_{n\neq 0} n c_{-n}\c _n\nonumber\\
d_R & = & \sum_{n\neq 0} c_{-n}[P^+(n)\a^-_n+P^-(n)\a^+_n] +
\sum_{\stackrel{n,m\neq 0}{n+m\neq 0}} :
c_{-n}[\a^+_{-m}\a^-_{m+n} +\frac{1}{2} (m-n)c_{-m}b_{m+n}]:\nonumber\\
& & -\frac{1}{2} \sum_{\stackrel{n\neq 0}{m}}(2m+n)c_{-n}
:\psi^+_{n+m}\psi^-_{-m}: -\frac{1}{2} \sum_{n\neq 0} \c_{-n}
[P^+(2n)\psi^-_n+P^-(2n)\psi^+_n] \nonumber\\
& & -\frac{1}{2} \sum_{n,m\neq 0}\c_{-m}(\psi^+_{m-n}\a^-_n
+\psi^-_{m-n}\a^+_n)\nonumber\\
& & + \sum_{\stackrel{n,m\neq 0}{n+m\neq 0}}
[(\frac{3}{2}n+m):c_{-n}\c_{n+m}\b_{-m}: -\frac{1}{4}
\c_n\c_mb_{-n-m}]
\ena
In the R sector, there is no subtraction in $L_0$~\cite{OHT,MIT}. Instead,
the $\frac{1}{2}$ coming from $p^+p^-$ is here absorbed in the normal
ordering of the zero modes, as discussed in ref.~\cite{ITO,TKU}.
The nilpotency of $Q_B$ is rewritten as
\bea
L_0 & = & F^2\;\;,\;\; d^2_R = M_RL_0 + KF\;\;,\;\; 2K = [M_R,
F]\nonumber\\
K^2 & = & [L_0, M_R] = [L_0, F]= [L_0, K] = [L_0, d_R]\nonumber\\
& = & [M_R, K] = [M_R, d_R] = \{F, K\} = \{F, d_R\} = \{K, d_R\} = 0
\ena

The Hilbert space ${\cal H}$ of the full theory is the direct sum
\EQ
{\cal H}= \oplus_{p^M,p^L}\left(
{\cal H}^{(M)}_{(p^M)} \otimes
{\cal H}^{(L)}_{(p^L)} \otimes {\cal H}^{(G)} \right)
\EN
where ${\cal H}^{(M)}_{(p^M)}$ (${\cal H}^{(L)}_{(p^L)}$) is the Fock
space of matter (Liouville) oscillators acting on a Fock vacuum with
momentum $p^M (p^L)$ and ${\cal H}^{(G)}$ is the ghost Hilbert space.

The physical state conditions in both sectors are given by
\EQ
Q_B \mid {\rm phys}> = 0.
\EN
Since in both cases $L_0 = \{b_0, Q_B\},$ these physical states satisfy
\EQ
L_0\mid{\rm phys}> = Q_Bb_0\mid{\rm phys}>.
\EN
Hence, any physical states are BRST-exact unless they satisfy the
on-shell condition $L_0 = 0.$

It is convenient to reduce the zero eigenspace of $L_0$ by restricting
to the states annihilated by $b_0$ (and also by $\b_0$ in the R sector).
In this space the physical state conditions (2.23) reduce to
\EQ
L_0\mid{\rm phys}>= b_0\mid {\rm phys}> = d_{NS} \mid{\rm phys}>=0
\EN
for the NS sector and to
\EQ
F\mid {\rm phys}> = b_0\mid {\rm phys}>=
\b_0\mid{\rm phys}>=d_R\mid{\rm phys}>=0
\EN
for the R sector. Note that the condition $L_0\mid{\rm phys}>=0$
for the R sector is satisfied due to the relation $L_0=F^2$.
Notice also $d^2=0$ when acting on this space because of the
relations (2.13) and (2.21).

\sect{Physical states in the NS sector}

In this section, we discuss the relative cohomology (2.25) for the
NS sector and identify physical states.

\subsection{Cohomology of $d_0$}

{}From the first condition $L_0\mid{\rm phys}>=0$ in (2.25) and
(2.18a), we see that this space is nontrivial only if $p^+p^-$
takes a non-positive half-integer or integer value. For $p^+p^-=0$,
there is a unique state $\mid p^M,p^L >$.

In order to examine the cohomology of $d_{NS}$, we introduce the
degree for the oscillators~\cite{KOG,BMP} as
\bea
&& deg\left(\a^+_n,\psi^+_n,c_n,\c_r\right) = +1\nonumber\\
&& deg\left(\a^-_n, \psi^-_n,b_n,\b_r\right) = -1
\ena
and define the degree of $\mid p^M,p^L>$ to be zero. The cohomology
operator $d_{NS}$ is then decomposed into components with definite
degrees:
\EQ
d_{NS} = d_0 +d_1+d_2.
\EN
Here
$$
d_0 = \sum_{n\neq 0} P^+(n)c_{-n}\a^-_n
-\frac{1}{2}\sum_rP^+(2r)\c_{-r}\psi^-_r, \eqno(3.3a)
$$
$$
d_1=\sum_{nm(n+m)\neq 0}:c_{-n}
\left[\a^+_{-m}\a^-_{m+n}+\frac{1}{2}(m-n)c_{-m}b_{m+n}\right]:
-\frac{1}{2} \sum_{\stackrel{n\neq 0}{r}}
\c_{-r}\left(\psi^+_{r-n}\a^-_n+\psi^-_{r-n}\a^+_n\right),
$$
$$
-\frac{1}{2}\sum_{\stackrel{n\neq 0}{r}}\left(2r+n\right):c_{-n}
\psi^+_{n+r}\psi^-_{-r}:+\sum_{\stackrel{n\neq 0}{r}}\left(
\frac{3}{2}n+r\right):c_{-n}\c_{n+r}\b_{-r}:
-\frac{1}{4}\sum_{r+s\neq 0}\c_r\c_sb_{-r-s}\eqno(3.3b)
$$
$$
d_2 = \sum_{n\neq 0} P^-(n)c_{-n}\a^+_n
-\frac{1}{2}\sum_rP^-(2r)\c_{-r}\psi^+_r\eqno(3.3c)
$$
satisfy in the on-shell subspace
\setcounter{equation}{3}
\EQ
d^2_0=d^2_2=0, \;\; \left\{d_0,d_1\right\}=\left\{d_1,
d_2\right\}=0,\;\; d^2_1+\left\{d_0,d_2\right\}=0.
\EN
Note that the argument in $P^+(2r)$ in the second term in (3.3a) is
an odd integer.

Our strategy is first to consider the nontrivial cohomology of $d_0$
and then examine if it can be extended to the cohomology of $d_{NS}$.
In the critical string~\cite{KOG,OHT,MIT} and the Liouville theory coupled
to matter~\cite{LIZ,BMP}, it has been shown that there is an isomorphism
between the nontrivial cohomology classes of $d_0$ and $d_{NS}$. In
our case of the super-Liouville coupled to superconformal matters, it
turns out that this is no longer true in general.  Nevertheless, we
will find it useful to examine the cohomology of $d_0$.

To compute the cohomology of $d_0$, we must consider
the following two cases:\\
I.  $P^+(n)\neq 0, P^-(n)\neq 0$ for all integers $n\neq 0$.\\
II. There exist integers $j,k$ such that $P^+(j)=P^-(k)=0.$

We will see later that if $P^+(n)$ or $P^-(n)$ vanishes for some
integer at all, the other must also vanish at some integer due to the
on-shell condition, and hence there is no other case than these two.
The case II is further devided into four possibilities: (i) even $j$
and odd $k$; (ii) even $j$ and $k$; (iii) odd $j$ and $k$;
(iv) odd $j$ and even $k$.

Let us examine the cohomology of $d_0$ in each case.

\underline{Case I. $P^+(n)\neq 0, P^-(n)\neq 0$ for all $n\neq 0$}

If we define
\EQ
K_{NS}\equiv \sum_{n\neq 0}\frac{1}{P^+(n)}\a^+_{-n}b_n
+\sum_r\frac{2r}{P^+(2r)}\psi^+_{-r}\b_r
\EN
the number operator $\hat N$ may be written as $\hat N=\{d_0,K_{NS}\}$.
This implies that any $d_0$-closed state of nonzero level is $d_0$-exact,
{\it i.e.} cohomologically trivial. Hence the only nontrivial cohomology
is obtained for $\hat N=0$, {\it i.e.}
\EQ
\mid p^M,p^L>\;\;\;{\rm with}\;\;\; p^+p^-=0.
\EN
which is the state we called the degree of freedom corresponding to
the ``center of mass".

\underline{Case II. $P^+(j) = P^-(k)= 0$}

Since $P^\pm (n)$ are linear in $n$, it follows from eq.~(2.14) that
\bea
P^+(n) & = & \frac{1}{\sqrt 2}\left(\la^M+i\la^L\right)(n-j)\nonumber\\
P^-(m) & = & \frac{1}{\sqrt 2}\left(\la^M-i\la^L\right)(m-k)
\ena
In particular, this implies that these are nonzero for other values of
$n$ and $m$. From (3.7) we see that
\EQ
p^+p^-=P^+(0)P^-(0)=\frac{1}{2}\left[(\la^M)^2 + (\la^L)^2\right] jk=
-\frac{1}{2} jk.
\EN
Combined with the on-shell condition, we find the level is given by
\EQ
\hat N = \frac{1}{2} jk\;\;\;.
\EN
Hence we have either $j,k>0$ or $j,k <0$.

\underline{(i) Even $j$ and odd $k$}

If we define
\EQ
K_j=\sum_{n\neq 0,j}\frac{1}{P^+(n)}\a^+_{-n}b_n
+\sum_r\frac{2r}{P^+(2r)}\psi^+_{-r}\b_r
\EN
then $\hat N_{0,j} = \{d_0, K_j\}$ is the level operator for all the
oscillators except $\a^+_{-j}$ and $c_{-j}$ ($\a^-_{j}$ and $b_{j}$)
when $j,k>0$ ($j,k<0$).  The cohomology of $d_0$ is thus constructed
from these mode operators. It turns out that no such state can satisfy
the on-shell condition. If $j=2(2l +1)$ for some integer $l$, the level
$\frac{1}{2} jk$ is an odd integer whereas all the available mode
operators have even levels. If $j=2\cdot 2l$, on the other hand, the
level is $2l k$ which cannot be made from the mode operators
$\a^+_{-4l}$ and $c_{-4l}$. Hence we conclude that the cohomology of
$d_0$ is trivial.

\underline{(ii) Even $j$ and $k$}

We can define the same level operator as in (i). The nontrivial
cohomology of $d_0$ is represented by the states
\EQ
\left(\a^+_{-j}\right)^{k/2}\mid p^M, p^L >\;\; ,\;\;
c_{-j}\left(\a^+_{-j}\right)^{k/2-1}\mid p^M, p^L >
\EN
for $j,k >0$ and
\EQ
\left(\a^-_j\right)^{-k/2}| p^M, p^L >\;\;,
b_j\left(\a^-_j\right)^{-k/2-1}| p^M, p^L >
\EN
for $j,k <0.$ By inspection, we see that these are indeed nontrivial
cohomology states. This is similar to the bosonic case~\cite{LIZ,BMP}.

\newpage
\underline{(iii) Odd $j$ and $k$}

In this case, we have to modify $K_j$ to
\EQ
K^\prime_j = \sum_{n\neq 0,j}\frac{1}{P^+(n)} \a^+_{-n}b_n +
\sum_{r\neq j/2}\frac{2r}{P^+(2r)}\psi^+_{-r}\b _r
\EN
and then $\hat N'_{0,j} = \left\{d_0, K'_j\right\}$ is the level
operator except for $\a^+_{-j},c_{-j}, \psi^+_{-j/2}$ and $\c_{-j/2}$
($\a^-_{j},b_{j}, \psi^-_{j/2}$ and $\b_{j/2}$) when $j,k>0$ $(j,k<0)$.
We find that the nontrivial cohomology of $d_0$ is represented by the
states listed below according to their degrees.

For $j,k>0$:
\EQ
\begin{array}{rrrr}
\mbox{degree} & & & \\
k: & &(\c_{-j/2})^k,\hspace{4cm} &\psi^+_{-j/2}(\c_{-j/2})^{k-1},
\hspace{1.5cm} \\
 & \nearrow & \nearrow & \\
k-1: &c_{-j}(\c_{-j/2})^{k-2},&\left\{
\begin{array}{ll}
\a^+_{-j}(\c_{-j/2})^{k-2}, \\
c_{-j} \psi^+_{-j/2}(\c_{-j/2})^{k-3},
\end{array} \right. \hspace{1cm}
&\psi^+_{-j/2}\a^+_{-j}(\c_{-j/2})^{k-3}\hspace{1cm}\\
 & \nearrow & \nearrow & \\
k-2: &c_{-j}\a^+_{-j}(\c_{-j/2})^{k-4},&\left\{
\begin{array}{ll}
(\a^+_{-j})^2(\c_{-j/2})^{k-4}, \\
c_{-j} \psi^+_{-j/2}\a^+_{-j}(\c_{-j/2})^{k-5},
\end{array} \right.
&\psi^+_{-j/2}(\a^+_{-j})^2(\c_{-j/2})^{k-5} \hspace{5mm}\\
 & \nearrow & \nearrow & \\
\cdots & \cdots\hspace{2cm}& \cdots\hspace{2cm}
& \cdots\hspace{2cm} \\
 & \nearrow & \nearrow & \\
\frac{k+3}{2}: &c_{-j}(\a^+_{-j})^{(k-5)/2}(\c_{-j/2})^3,
&\left\{
\begin{array}{ll}
(\a^+_{-j})^{(k-3)/2}(\c_{-j/2})^3, \\
c_{-j} \psi^+_{-j/2}(\a^+_{-j})^{(k-5)/2}(\c_{-j/2})^2,
\end{array} \right.
&\psi^+_{-j/2}(\a^+_{-j})^{(k-3)/2}(\c_{-j/2})^2 \\
 & \nearrow & \nearrow & \\
\frac{k+1}{2}: &c_{-j}(\a^+_{-j})^{(k-3)/2}\c_{-j/2},
 &\left\{
\begin{array}{ll}
(\a^+_{-j})^{(k-1)/2}\c_{-j/2}, \\
c_{-j} \psi^+_{-j/2}(\a^+_{-j})^{(k-3)/2},
\end{array} \right.
&\psi^+_{-j/2}(\a^+_{-j})^{(k-1)/2} \hspace{1cm}
\end{array}
\EN

For $j,k<0$:
\EQ
\begin{array}{llll}
k:& &(\b_{j/2})^{-k}, &\psi^-_{j/2}(\b_{j/2})^{-k-1},\\
 & & \swarrow & \swarrow \\
k+1: &\hspace{1.5cm}b_{j}(\b_{j/2})^{-k-2},&\hspace{5mm}\left\{
\begin{array}{ll}
\a^-_{j}(\b_{j/2})^{-k-2}, \\
b_{j} \psi^-_{j/2}(\b_{j/2})^{-k-3},
\end{array} \right.
&\psi^-_{j/2}\a^-_{j}(\b_{j/2})^{-k-3}\\
 & & \swarrow & \swarrow \\
\cdots & \hspace{2cm}\cdots & \hspace{2cm}\cdots
& \hspace{2cm}\cdots \\
 & & \swarrow & \swarrow \\
\frac{k-3}{2}: &b_{j}(\a^-_{j})^{-(k+5)/2}(\b_{j/2})^3,
&\left\{
\begin{array}{ll}
(\a^-_{j})^{-(k+3)/2}(\b_{j/2})^3, \\
b_{j} \psi^-_{j/2}(\a^-_{j})^{-(k+5)/2}(\b_{j/2})^2,
\end{array} \right.
&\psi^-_{j/2}(\a^-_{j})^{-(k+3)/2}(\b_{j/2})^2 \\
 & &\swarrow & \swarrow \\
\frac{k-1}{2}: &\hspace{5mm}b_{j}(\a^-_{j})^{-(k+3)/2}\b_{j/2},
 &\hspace{1cm} \left\{
\begin{array}{ll}
(\a^-_{j})^{-(k+1)/2}\b_{j/2}, \\
b_{j} \psi^-_{j/2}(\a^-_{j})^{-(k+3)/2},
\end{array} \right.
&\psi^-_{j/2}(\a^-_{j})^{-(k+1)/2}
\end{array}
\EN

Here the ground state $|p^M,p^L>$ with $p^+p^- = -\shalf jk$ is not
exposed explicitly and the arrows indicate flows under the action of
$d_{NS}$ to be discussed in the next subsection. We thus see that there
are many states with various ghost numbers.

\underline{(iv) Odd $j$ and even $k$}

The number operator is the same as in (iii). The states to span the
nontrivial space are, however, a little different from the case (iii)
in the bottom parts of the tables:

For $j,k>0$:
\EQ
\begin{array}{rrrr}
k: & &(\c_{-j/2})^k,\hspace{4cm} &\psi^+_{-j/2}(\c_{-j/2})^{k-1},
\hspace{1cm} \\
 & \nearrow & \nearrow & \\
k-1: &c_{-j}(\c_{-j/2})^{k-2},&\left\{
\begin{array}{ll}
\a^+_{-j}(\c_{-j/2})^{k-2}, \\
c_{-j} \psi^+_{-j/2}(\c_{-j/2})^{k-3},
\end{array} \right. \hspace{1cm}
&\psi^+_{-j/2}\a^+_{-j}(\c_{-j/2})^{k-3}\hspace{5mm}\\
 & \nearrow & \nearrow & \\
\cdots & \cdots\hspace{2cm} & \cdots\hspace{2cm}
& \cdots\hspace{2cm} \\
 & \nearrow & \nearrow & \\
\frac{k+2}{2}: &c_{-j}(\a^+_{-j})^{(k-4)/2}(\c_{-j/2})^2,
&\left\{
\begin{array}{ll}
(\a^+_{-j})^{(k-2)/2}(\c_{-j/2})^2, \\
c_{-j} \psi^+_{-j/2}(\a^+_{-j})^{(k-4)/2}\c_{-j/2},
\end{array} \right.
&\psi^+_{-j/2}(\a^+_{-j})^{(k-2)/2}\c_{-j/2}\\
 & \nearrow & \nearrow & \\
\frac{k}{2}: &c_{-j}(\a^+_{-j})^{(k-2)/2},
 &(\a^+_{-j})^{k/2}&
\end{array}
\EN

For $j,k<0$:
\EQ
\begin{array}{llll}
k:& &(\b_{j/2})^{-k}, &\psi^-_{j/2}(\b_{j/2})^{-k-1},\\
 & & \swarrow & \swarrow \\
k+1: &\hspace{1cm}b_{j}(\b_{j/2})^{-k-2},&\hspace{5mm}\left\{
\begin{array}{ll}
\a^-_{j}(\b_{j/2})^{-k-2}, \\
b_{j} \psi^-_{j/2}(\b_{j/2})^{-k-3},
\end{array} \right.
&\psi^-_{j/2}\a^-_{j}(\b_{j/2})^{-k-3}\\
 & & \swarrow & \swarrow \\
\cdots & \hspace{2cm}\cdots & \hspace{2cm}\cdots
& \hspace{2cm}\cdots \\
 & & \swarrow & \swarrow \\
\frac{k-2}{2}: &b_{j}(\a^-_{j})^{-(k+4)/2}(\b_{j/2})^2,
&\left\{
\begin{array}{ll}
(\a^-_{j})^{-(k+2)/2}(\b_{j/2})^2, \\
b_{j} \psi^-_{j/2}(\a^-_{j})^{-(k+4)/2}\b_{j/2},
\end{array} \right.
&\psi^-_{j/2}(\a^-_{j})^{-(k+2)/2}\b_{j/2}\\
 & & \swarrow & \swarrow \\
\frac{k}{2}: &\hspace{1cm}b_{j}(\a^-_{j})^{-(k+2)/2},
 & \hspace{3cm}(\a^-_{j})^{-k/2},&
\end{array}
\EN

Here again the ground state is suppressed.

Finally let us show that we have exhausted all possible cases. Suppose
that $P^+(j)=0$ but $P^-(n) \neq 0$ for all nonzero integers $n$. Owing
to the linearity in $n$, $P^+(n)$ is rewritten like eq.~(3.7)
\EQ
P^-(m) = \frac{1}{\sqrt 2} \left(\la^M-i\la^L\right) (m-\a)
\EN
but with $\a$ being not an integer. This gives us
\EQ
p^+p^- = P^+(0)P^-(0)= -\frac{1}{2} j \a
\EN
which implies that the nontrivial cohomology is possible at the level
$\frac{1}{2}j\a$. However, as is evident from the above analysis,
the only available mode operators for creating cohomologically
nontrivial states have the level either $j$ or $j/2$, which cannot
give the level $\shalf j\a$. Therefore there is no nontrivial
cohomology of $d_0$ in this case. The same argument excludes the case
when $P^-(k)=0$ but $P^+(n)\neq 0$.

\subsection{Cohomology of $d_{NS}$}

In order to construct a state representing nontrivial cohomology of
$d$, \footnote{We suppress the subscript $NS$ to $d$ in what follows
since most of the following discussions are valid for $d_R$ as well.}
we may start from a state nontrivial with respect to $d_0$ and add terms
of higher degrees.  This is due to the following fact. The lowest degree
term in a state nontrivial with respect to $d$ may be always chosen to
represent a nontrivial cohomology of $d_0$; or if the cohomology of
$d_0$ is trivial, then the cohomology of $d$ is also trivial.

The proof is simple~\cite{BMP}. Suppose that $\psi =\psi_k +\psi_{k+1}
+\cdots$ represents a cohomology of $d$, where the subscripts $k, k+1,
\cdots$ stand for the degrees. Then $d\psi =0$ means $d_0\psi_k=0$ and
thus $\psi_k=d_0\chi_k$ by assumption. If we consider $\psi^\prime=\psi
-d\chi_k$ which also belongs to the same cohomology class as $\psi$,
the lowest degree term in $\psi '$ has degree at least $k+1$. Repeating
the same procedure as above, we arrive at $\psi=d(\chi_k+\chi_{k+1}
+\cdots)$, which is the desired result.

Furthermore, if for each ghost number $N_{FP}$, the cohomology of $d_0$
is nontrivial for at most one fixed degree $k$ independent of $N_{FP}$,
then the nontrivial cohomology classes of $d_0$ and $d$ are isomorphic.
We only sketch the proof briefly. (See ref.~\cite{BMP} for the details.)

Let $\psi_k$ be a nontrivial element in the cohomology of $d_0$. Then
$d\psi_k=(d_1+d_2)\psi_k$ has the lowest degree at least $k+1$. Using
(3.4), we find $d_0(d_1\psi_k)=0$. Since there is no nontrivial
cohomology at degree $k+1$, we get $d_1\psi_k=d_0\chi_{k+1}$. Then
$d(\psi_k-\chi_{k+1})$ has terms of degree at least $k+2,$ and the use
of eq.~(3.4) tells us that $d_0[d(\psi_k-\chi_{k+1})]_{k+2}=0$.
Repeating the above procedure, we construct $\psi\equiv
\psi_k-\chi_{k+1}-\chi_{k+2}-\cdots$ such that $\psi$ is closed under
$d$. Moreover, one can show that this map gives a unique element in
$d$-cohomology. Conversely, it can be shown that each element of
cohomology of $d$ projects onto a unique element of $d_0$-cohomology.
This completes the proof.

Since the assumption here is satisfied for cases I and II (i) and (ii),
we find nontrivial cohomology of $d$ only for ${\hat N}=0$ and $\shalf
jk$ for even $j$ and $k$ with ghost number $N_{FP}=0,\pm 1.$

In the other cases, the above statement does not apply because there
are many nontrivial cohomology classes of $d_0$. In fact, when $d$ acts
on the states in the tables (3.14)--(3.17), there appear other states
of the nontrivial cohomology of $d_0$ as indicated by the arrows
in (3.14)--(3.17). For example, for the case (iii) the states transform
under the action of $d$ as
\bea
&& c_{-j}(\c_{-j/2})^{k-2}| p^M,p^L >
\to (\c_{-j/2})^k | p^M, p^L > \to 0\nonumber\\
&& c_{-j}\a^+_{-j}(\c_{-j/2})^{k-4}| p^M, p^L > \to
[-\frac{1}{4} \a^+_{-j}(\c_{-j/2})^{k-2}+\frac{j}{2}
c_{-j}\psi^+_{-j/2}(\c_{-j/2})^{k-3}]| p^M, p^L >\nonumber\\
&& \to 0
\ena
where we have written only the terms nontrivial in the $d_0$-cohomology.
It is important to note that these states always vanish under the second
action of $d$ and that states nontrivial with respect to $d_0$ are
produced only at the next degree and ghost number. The first fact is a
reflection of the nilpotency of the BRST charge and the second is due
to the fact that the states are created by the action of $d_1.$ If we
call the initial states ``parents" and the resulting states
``daughters", we can check that most of these states in the tables are
either parents or daughters.

Now we prove that neither parents nor daughters can give rise to
nontrivial cohomology of the total $d$ even if higher degree
terms are added.

To show that the parents do not produce any, assume that we succeed
in constructing a state $\psi=\psi_k+\psi_{k+1}+\cdots$ such that
\EQ
d(\psi_k+\psi_{k+1}+\cdots)=0
\EN
starting from a parent $\psi_k.$ Since $d_0\psi_k=0$, the lowest
degree terms in (3.21) give
\EQ
d_1\psi_k+d_0\psi_{k+1}=0.
\EN
According to (3.4), $d_0(d_1\psi_k) = -d_1d_0\psi_k
= 0$ and thus we get
\EQ
d_1\psi_k = \eta_{k+1}+d_0\chi_{k+1}
\EN
where $\eta_{k+1}$ denote a state nontrivial with respect to $d_0$.
(From parents, states corresponding to nontrivial cohomology classes
of $d_0$ are always produced by the action of $d_1$.) But if we
substitute (3.23) into (3.22), we obtain
\EQ
\eta_{k+1}=-d_0(\chi_{k+1}+\psi_{k+1})
\EN
in contradiction to the fact that $\eta_{k+1}$ is nontrivial.
Therefore the parents are excluded.

Let us next consider a daughter state, $\eta_{k+1}$, obtained from a
parent state, $\psi_k$, as given in (3.23). This relation can be
rewritten as
\EQ
d(\psi_k - \chi_{k+1})
= \eta_{k+1} + [-d_1 \chi_{k+1} +d_2 \psi_k] - d_2 \chi_{k+1}
\EN
Thus we obtain a $d$-trivial state by adding degree $k+2$ and $k+3$
terms to $\eta_{k+1}$.  Let us write this as $\eta_{k+1} + \eta_>$.
Suppose that there is a $d$-nontrivial state with $\eta_{k+1}$ as the
lowest degree term and write it as $\eta_{k+1} + \eta'_>$. We may take
a representative of the same cohomology class as
\EQ
\eta_{k+1} + \eta'_> - d(\psi_k - \chi_{k+1})
\EN
However this does not contain $\eta_{k+1}$. Thus there is no nontrivial
cohomology class represented by a state with $\eta_{k+1}$ as the
lowest degree term. Hence the above statement is proved.

Using the above results, we can discard all the parents and daughters
for constructing the cohomology of $d$. In particular, for case II (iv)
all the states are either parents or daughters. We find that the only
exceptions are
\bea
\psi^+_{-j/2}(\a^+_{-j})^{(k-1)/2}| p^M,p^L>, \nonumber\\
{} [(\a^+_{-j})^{(k-1)/2}\c_{-j/2}-
j(k-1)c_{-j}\psi ^+_{-j/2}(\a ^+_{-j})^{(k-3)/2}]| p^M,p^L>
\ena
for odd $j,k >0$ and
\bea
\psi^-_{j/2}(\a^-_{j})^{-(k+1)/2}| p^M,p^L>, \nonumber\\
{} [(\a^-_{j})^{-(k+1)/2}\b_{j/2}-
\shalf b_{j}\psi^-_{j/2}(\a^-_{j})^{-(k+3)/2}]| p^M,p^L>
\ena
for odd $j,k <0.$ The linear combinations are singled out by the
requirement that they are neither parents nor daughters.

It can be shown that these states can be promoted to the cohomology of
$d$ by adding higher degree terms by a procedure described before for
$d_0$-nontrivial states. It is also easy to see that $\psi_k-\chi_{k+1}
-\chi_{k+2}-\cdots$ thus constructed is not trivial. (If we assume that
it is trivial, it leads to a contradiction that $\psi_k$ is
$d_0$-trivial.)

To summarize, we have found that there are nontrivial states for
$p^+p^-=0$ and for $p^+p^-=-\shalf jk$ with $j-k$= even. In the latter
case, the states can be constructed from those in eqs.~(3.11) and (3.12)
for even $j$ and $k$, and from those in eqs.~(3.27) and (3.28) for odd
$j$ and $k$.

\subsection{Quartet mechanism}

In the previous subsections we have examined the nontrivial cohomology
classes for $d_{NS}$. We have encountered quite a different situation
from the bosonic case where there is a one-to-one correspondence
between the cohomologies of $d_0$ and $d_{NS}$, and yet have succeeded
in identifying the nontrivial classes of $d_{NS}$. Here we will give an
alternative derivation of this result by showing that the states, which
are found to be $d$-trivial but $d_0$-nontrivial in the previous
subsection, actually fall into the so-called BRST quartet
representations and hence decouple from the system~\cite{KUO}. Although
the final results are the same as in the previous subsection, we
believe that this reformulation shows the essence of the decoupling
mechanism of the states and also it is more accessible to physicists.
Since the other cases are essentially the same, let us consider the NS
sector with both $j$ and $k$ odd integers.

The Fock vacuum is the direct product of the vacua for the matter,
Liouville and ghost systems
\EQ
|p^M, p^L> \equiv |p^M> \otimes |p^L> \otimes |0>_{gh}.
\EN
{}From the conditions $P^{+}(j)=P^{-}(k)=0$, we find the momenta are
given as
\bea
p^{M,L} = \shalf (jt^{M,L}_+ + kt^{M,L}_-)
=t^{M,L}_{(j,k)}+\la^{M,L}
\ena
where $t^M_{\pm}=-\la^M \mp i\la^L,~t^L_{\pm}=-\la^L \pm i\la^M$ and
$t^{M,L}_{(j,k)} \equiv \frac{1+j}{2}t^{M,L}_+ +\frac{1+k}{2}
t^{M,L}_-$. Thus the vacuum in (3.29) may also be labelled by
these integers as $|(j,k)>$.

Let us first observe the following pattern in the $d_0$-nontrivial
states in (3.14) and (3.15).

For the momenta satisfying $P^{+}(j)=P^{-}(k)=0$ with $j, k>0$
\EQ
\begin{array}{c|c|c|c}
\begin{array}{c} N_{FP} \\  \mbox{deg.}  \end{array}
&       k-(2l+1)        &  k-(2l+2)     &  k-(2l+3)     \\ \hline
k-l     & |k-(2l+1)>    &       &       \\ \hline
k-l-1   & **    &
\begin{array}{c}
|k-(2l+2)>_+ \\
|k-(2l+2)>_- \end{array} & ** \\ \hline
k-l-2   &       &       & |k-(2l+3)>
\end{array}
\EN
and for $P^{+}(-j)=P^{-}(-k)=0$ with $j,k>0$
\EQ
\begin{array}{c|c|c|c}
\begin{array}{c} N_{FP} \\  \mbox{deg.}  \end{array}
&       -k+(2l+1)       &  -k+(2l+2)    &  -k+(2l+3)    \\ \hline
-k+l    & |-k+(2l+1)>   &       &       \\ \hline
-k+l+1  & **    &
\begin{array}{c}
|-k+(2l+2)>_+ \\
|-k+(2l+2)>_- \end{array} & ** \\ \hline
-k+l+2  &       &       & |-k+(2l+3)>
\end{array}
\EN
Here we have defined the states by
\newpage
\bea
|k-(2l+1)>
&\equiv & \frac{l+1}{4}j(\a^+_{-j})^{l}\psi^+_{-j/2}
(\c_{-j/2})^{k-(2l+1)}|(j,k)>, \nonumber\\
|k-(2l+2)>_{\pm}
&\equiv & [ -\frac{1}{4}(\a^+_{-j})^{l+1}(\c_{-j/2})^{k-(2l+2)}
\nonumber\\
&&   \pm \frac{l+1}{2}jc_{-j}(\a^+_{-j})^l\psi^+_{-j/2}
    (\c_{-j/2})^{k-(2l+3)}] |(j,k)>,  \nonumber \\
|k-(2l+3)>
&\equiv& c_{-j}(\a^+_{-j})^{l+1}(\c_{-j/2})^{k-(2l+4)}|(j,k)>,
\nonumber \\
|-k+(2l+1)>
&\equiv & \frac{(-1)^{l+1}4}{j^{l+1}(l+1)!(k-2l-1)!}
(\a^-_{-j})^l\psi^-_{-j/2}(\b_{-j/2})^{k-(2l+1)}|(-j,-k)>,
\nonumber\\
|-k+(2l+2)>_{\pm}
&\equiv & \frac{(-1)^{l+1}}{j^{l+1}(l+1)!(k-2l-2)!}
  [- 2(\a^-_{-j})^{l+1}(\b_{-j/2})^{k-(2l+2)}
\nonumber\\
&&  \mp (k-2l-2)b_{-j}(\a^-_{-j})^l\psi^-_{-j/2}
    (\b_{-j/2})^{k-(2l+3)}] |(-j,-k)>,  \nonumber\\
|-k+(2l+3)> &\equiv& \frac{(-1)^{l+1}}{j^{l+1}(l+1)!(k-2l-4)!} \nonumber\\
&&\times b_{-j}(\a^-_{-j})^{l+1}(\b_{-j/2})^{k-(2l+4)}|(-j,-k)>.
\ena
In the tables $l$ runs from $-1$ to $\frac{k-3}{2}$ and the double
asterisks mean that there are other states for adjacent values of
$l$. (It is to be understood that when the power of the mode operators
becomes negative, there is no corresponding state.)

The proper inner product is to be defined, in an abbreviated
form\footnote{As discussed in ref.~\cite{IT2}, there is another
possibility to define an inner product when ${\hat c}=1$. Here
we adopt the convention defined in the text.}, as
\EQ
<O, O'> \equiv (O|-p^M,-p^L>)^{\dagger} P c_0 O'|p^M,p^L>
\EN
where $P$, a parity operator~\cite{KMA}, changes the signs of the
oscillators including zero modes in the matter sector. It is important
to note that the inner product in (3.34) respects the hermiticity of
the Virasoro generators
\EQ
P L_n^\dagger (\la^{M,L}) P = L_{-n} (\la^{M,L}).
\EN

In the following, we will show that there are two sets of quartets
associated with the states listed in eqs.~(3.31) and (3.32).

As we have already seen in simple examples in eq.(3.20), the states are
transformed under the action of $d_{NS}$ as
\bea
d_{NS}|k-(2l+3)> &=& |k-(2l+2)>_+ +d_0\chi^{k-(2l+2)} \nonumber\\
 &\equiv& |k-(2l+2)>_+', \nonumber\\
d_{NS}|k-(2l+2)>_- &=& |k-(2l+1)> +d_0\chi^{k-(2l+1)}\nonumber\\
 &\equiv& |k-(2l+1)>'.
\ena
Note that the terms other than $d_0$-nontrivial states may be written
in $d_0$-exact forms, denoted as $d_0\chi^{k-(2l+2)}$ and
$d_0\chi^{k-(2l+1)}$. These parent and daughter states in (3.36)
form the doublet representations of the BRST charge.
For the states in eq.~(3.32), we find similar relations
\bea
d_{NS}|-k+(2l+1)> &=& |-k+(2l+2)>_+ +d_0\chi^{-k+2l+2}, \nonumber\\
 &\equiv& |-k+(2l+2)>_+', \nonumber\\
d_{NS}|-k+(2l+2)>_- &=& |-k+(2l+3)> +d_0\chi^{-k+2l+3} \nonumber\\
 &\equiv & |-k+(2l+3)>'.
\ena

For the following discussion, it is important to realize that
the states in eq.~(3.33) have nonvanishing inner products
\EQ
-< -k+(2l+3)|'c_0|k-(2l+3)>=
_-< -k+(2l+2)|c_0|k-(2l+2)>'_+=1.
\EN
where we have used the relation $P(\a_n^{\pm})^\dagger P = - \a_{-n}^
{\pm}$ and $P(\psi_n^{\pm})^\dagger P=-(\psi_{-n}^{\pm})$. Note also
that the $d_0$-exact terms do not contribute to the inner products.

{}From eqs.~(3.36)--(3.38), we see that those four states in (3.38) form
a BRST quartet, a pair of doublets with respect to $d_{NS}$. The quartet
mechanism first considered in nonabelian gauge theories then tells us
that these states decouple from the system~\cite{KUO}. To see this, it
is enough to note that the projection operator to the states in the
quartets in eq.~(3.38) is given by
\EQ
{\cal P} \equiv
\{d_{NS}, |k-(2l+3)>\cdot_-<-k+(2l+2)|c_0
-|-k+(2l+2)>_-\cdot<k-(2l+3)|c_0 \}
\EN
due to eqs.~(3.36)--(3.38). It follows that the quartet appears only in
$d_{NS}$-exact forms and all of the four states decouple from the
system.  One may repeat the same argument for the other states in
(3.36) and (3.37).

It is not difficult to see that most of the states in the tables
(3.14)--(3.17) fall into quartet representations as illustrated above.
The only exceptions are the states in (3.27) and (3.28), which
therefore contribute to the physical spectrum.

\subsection{Absolute cohomology}

Having identified the relative cohomology, let us make brief comments
on the absolute cohomology specified by (2.23).

The difference between (2.23) and the relative cohomology (2.25) is that
in the latter case we choose a special vacuum $|\downarrow \;>$
annihilated by $b_0$. Hence the absolute cohomology is obtained
essentially by adding the states built on the other vacuum
$|\uparrow \;>\equiv c_0|\downarrow \;>$. Indeed, it is not difficult,
following the bosonic case~\cite{LIZ,BMP}, to show that the absolute
cohomology is isomorphic to the direct product of these two relative
cohomologies.

\sect{Physical states in the R sector}

In this section, we proceed to the discussion of the relative
cohomology (2.26) in the R sector.

Let us first consider the subspace $V_F$ defined by the condition $F=0$.
The Ramond-Dirac operator $F$ in eq.(2.20) contains the zero modes
$\psi^{\pm}_0$ which are actually two-dimensional gamma matrices. We use
the following representation:
\EQ
\psi^+_0 = \left(
\begin{array}{cc}
0       & 1 \\
0       & 0 \end{array} \right), \hspace{5mm}
\psi^-_0 = \left(
\begin{array}{cc}
0       & 0 \\
1       & 0 \end{array} \right).
\EN
It is then convenient to understand that ${\hat F}$, the nonzero-mode
part of $F$, is multiplied by $\sigma_3$ so that it automatically
anticommutes with (4.1). Any spinor in this representation can be
written as
\EQ
\left( \begin{array}{c}
|A> \\ |B>  \end{array} \right)
\EN
where $|A>$ and $|B>$ denote the states spanned by the mode operators
and carrying momenta.

On this state the condition $F|{\rm phys}>=0$ becomes
\EQ
\left( \begin{array}{c}
p^-|B>+{\hat F}|A> \\
p^+|A>-{\hat F}|B> \end{array} \right) =0.
\EN
The minus sign in the lower column is due to $\sigma_3$.
Since $F^2=L_0 = p^+p^- +\hat F^2$, eq. (4.3) may be solved as follows.
When $p^-\neq 0,$ we take an on-shell state $|A>$ ($L_0|A>=0$) and
define $|B> = -\frac{1}{p^-}\hat F|A>$. Then we have
\EQ
{\hat F}|B>=-\frac{1}{p^-}{\hat F}^2|A>=p^+|A>.
\EN
Thus the condition (4.3) is satisfied. If $p^+\neq 0$, we can similarly
prepare the state $| B>$ and construct a spinor satisfying (4.3).
Therefore we can always construct $F=0$ spinors from states satisfying
$L_0=0$. When $p^{\pm}=0$, $p^{M,L}=0$ and no oscillators can be excited
owing to the on-shell condition $L_0=0$. The solution of the condition
(4.3) is given by a constant spinor mutiplied by $|p^M=0> \otimes
|p^L=0> \otimes |0>_{gh}$. This spinor is already a solution of $d_R=0$.
So we may consider cases of $p^+ \ne 0$ or $p^- \ne 0$ in the following.

Next we examine the condition $d_R=0$ in $V_F$. If decompose $d_R$
according to the degrees (the degrees of the zero modes are zero), we
find terms with zero modes in $d_0$. This is a slightly different
situation from NS sector. However we may recover the similarity by
introducing the following mode operators:
\bea
{\tilde \a}_n^\pm &\equiv& \a_n^\pm + n\theta\psi_n^\pm,
 \hspace{1cm}
{\tilde \psi}_n^\pm \equiv \psi_n^\pm - \theta\a_n^\pm
 \nonumber\\
{\tilde c}_n &\equiv& c_n-\theta\c_n, \hspace{1cm}
{\tilde b}_n \equiv b_n+n\theta\b_n \nonumber\\
{\tilde \c}_n &\equiv& \c_n+n\theta c_n, \hspace{1cm}
{\tilde \b}_n \equiv \b_n-\theta b_n,
\ena
where $\theta = \psi_0^+/p^+$, ($\theta^2=0$) for $p^+\neq 0$. The
commutation relations among these operators are the same as the original
ones without tildes. In terms of these operators, $d_R$ is decomposed as
\EQ
d_R=d_0+d_1+d_2
\EN
with
\EQ
d_0=\sum_{n \neq 0} P^+(n){\tilde c}_{-n}{\tilde \a}^-_n
  -\shalf \sum_{n \neq 0}P^+(2n) {\tilde c}_{-n} {\tilde \psi}^-_n.
\EN
This is quite similar to $d_0$ for the NS sector, although $d_0$
in the R sector is a matrix due to the zero mode dependence.
Another important difference from the NS sector is that the argument of
$P^+(2n)$ is an even integer. The operators $d_{0,1,2}$ satisfy (3.4)
and anticommute with $F$. The latter property is easily obtained from
(2.21) and deg($F$)$=0$. The modification of the oscillators is closely
related to that introduced in \cite{MIT} and used in \cite{ITO} to
solve the condition $F=0$. The above form was also suggested in
\cite{BMP2}.

We again have two different cases I and II in sect.~3. Let us first
enumerate the nontrivial cohomology of $d_0$.

\underline{Case I. $P^+(n)\neq 0,P^-(n)\neq 0$ for all $n\neq 0$}

In this case we may define
\EQ
K_R \equiv \sum_{n \neq 0} \frac{1}{P^+(n)}{\tilde \a}^+_{-n}{\tilde b}_n+
  \sum_{n \neq 0} \frac{2n}{P^+(2n)}{\tilde \psi}^+_{-n}{\tilde \b}_n
\EN
and the number operator $\hat N$ for modified oscillators is given as
$\hat N = \{d_0, K_R\}.$ Hence the nontrivial states do not have any
oscillator excitations and have $p^+=0$ or $p^-=0$ owing to the on-shell
condition:
\EQ
\left( \begin{array}{c}
0 \\ 1  \end{array} \right)\cdot
|p^M, p^L> \hspace{5mm} \mbox{for}\hspace{5mm} p^+=0, \hspace{1cm}
\left( \begin{array}{c}
1 \\ 0  \end{array} \right)\cdot
|p^M, p^L> \hspace{5mm} \mbox{for}\hspace{5mm} p^-=0.
\EN

\underline{Case II. $P^+(j)=P^-(k) =0$}

Since $p^+p^-=-\frac{1}{2}jk$ as in (3.8), we have the level
${\hat N}=\shalf jk >0$ from the on-shell condition $L_0=0$, and again
we have either $j,k>0 $ or $j,k<0.$  Obviously $p^{\pm} \ne 0$ and
oscillators in (4.5) are well-defined.

\underline{(i) Even $j$ and odd $k$}

Similarly to case (iii) in the NS sector, we should define
\EQ
K'_j \equiv \sum_{n \neq 0,j} \frac{1}{P^+(n)}{\tilde \a}^+_{-n}
{\tilde b}_n+
  \sum_{n \neq 0,j/2} \frac{2n}{P^+(2n)}
  {\tilde \psi}^+_{-n}{\tilde \b}_n
\EN
and the nontrivial cohomology of $d_0$ is given precisely by the
states in (3.14) and (3.15) with obvious replacement of oscillators
by the modified ones. The appropriate vacuum state is a spinor, whose
form can be read off in the following way. The vacuum does not depend
on the details of the various oscillator excitations in (3.14) and
(3.15), as we show now. Let us take the simplest example
$({\tilde \gamma}_{-j/2})^k$ in (3.14) for $p^+\neq 0$ and write it as
\EQ
({\tilde \gamma}_{-j/2} e^{-ijq^+/2p^+})^k e^{ijkq^+/2p^+}|\rho>
\equiv
({\tilde \gamma}_{-j/2})^k e^{-ikjq^+/2p^+}|\rho'>.
\EN
The oscillators with exponential factor commute with F \cite{MIT}, so
the condition $F=0$ gives $F|\rho'>=0$. The state $|\rho'>$ should
have level ${\hat N}=0$ and hence $p^+{}'p^-{}'=0$ from $L_0=0$. The solution
for $|\rho'>$ is given by the second state in (4.9). Since the momentum
$p^-=-jk/2p^+$ is added by the exponent in (4.11), the state (4.11)
gives the desired nontrivial spinor with momentum $p^+p^-=-jk/2$. The
same vacuum spinor should be used for all the states in (3.14) once the
momenta are specified. One may understand $p^-\neq 0$ case similarly.
The above consideration also applies to the cases listed below.

\underline{(ii) Even $j$ and $k$}

We can use the same level operator as in (i). The nontrivial cohomology
of $d_0$ is given by the states in (3.16) and (3.17) with modifications
described above.

\underline{(iii) Odd $j$ and $k$}

In this case, we can define
\EQ
K_j \equiv \sum_{n \neq 0,j} \frac{1}{P^+(n)}{\tilde \a}^+_{-n}
{\tilde b}_n+
  \sum_{n \neq 0} \frac{2n}{P^+(2n)}
  {\tilde \psi}^+_{-n}{\tilde \b}_n
\EN
and the nontrivial cohomology is possible in terms of ${\tilde \a}^+_{-j},
{\tilde c}_{-j}$ (${\tilde \a}^-_{j}, {\tilde b}_{j}$) for $j,k>0 (j,k<0)$.
However, we cannot construct states with the level $\frac{1}{2}jk$
(half-odd-integer) from integer mode operators.  Thus there is no
nontrivial cohomology in this case.

\underline{(iv) Odd $j$ and even $k$}

The number operator is the same as in (iii) above and the nontrivial
cohomology of $d_0$ is given by the space spanned by the states (3.11)
and (3.12).

\vspace{5mm}
Again here is no other possible case that could give rise to nontrivial
cohomology in the R sector.  The argument is the same as in the NS case
(cf. (3.19)).

As proved in sect.~3, parents and daughters under the action of $d_R$
do not correspond to nontrivial cohomology of $d_R$. By studying how
these states transform into each other, we find that the nontrivial
cohomology of $d_R$ is possible only for $j-k=$ odd in case II. For
even $j$ and odd $k$, it is constructed from the states given in
eqs.~(3.27) and (3.28). For odd $j$ and even $k$, the states are
obtained from (3.11) and (3.12).

We thus again find that nontrivial cohomology classes are possible only
at levels where ``null states" in the minimal models with $\hat c < 1$
exist~\cite{KMA}. We will discuss why this is so in the next section.

The absolute cohomology is again obtained by considering the vacua of
the ghost zero modes. For the $b-c$ ghost, this is essentially the
same as NS case. The vacua for the bosonic ghosts $\b -\c$ are
infinitely degenerate and we expect this leads to infinite number
of such spaces corresponding to these degrees of freedom.

\sect{Discussions}

Using the cohomological terms, we have examined the nontrivial states
allowed in the physical state conditions (2.25) and (2.26) for all the
cases in the NS and R sectors. Remarkably we have found, apart from the
ground state $|p^M, p^L>$ with $p^+p^-=0$, there exist nontrivial
states at levels ${\hat N}=\shalf jk$ and the discrete values of
momenta (3.30) with $p^+p^-=-\shalf jk$ . These states exist only for
$j-k=$ even (odd) in the NS (R) sector.

These are precisely the values of momenta at which special states with
respect to the Virasoro algebras appear in the Fock spaces of free
fields with background charges~\cite{KMA,FF}. Let us discuss why this
happens. We consider the bosonic case for simplicity of presentation
since the structure of the cohomology states is essentially the same
for the bosonic and supersymmetric cases as we have seen in this paper.
We must remember that in the bosonic case the extra states are present
at levels $N=jk$.

Let us first recall some properties of free field realization of a
conformal field theory (the matter or gravity sector). An important
quantity for our argument is the $C$-matrix considered in ref.
\cite{KMA}, relating the complete sets of states spanned by the Virasoro
generators and the boson oscillators. At level $N$ it is defined by
\EQ
L^{-I}(\la)|t+\la>=\sum_J C_{IJ}( p, \la)\a^{-J}|t+\la>
\EN
where $|t+\la>$ is a Fock vacuum with the momentum $t+\la$, and $L^{-I}
(\la)$ and $\a^{-J}$ stand for all the independent combinations of the
Virasoro generators and oscillators at level $N$, respectively. Thus
$I$ and $J$ run from 1 to $P(N)$, the partition number of the integer
$N$. It has been shown in ref.~\cite{KMA} that the $C$-matrix satisfies
$$
det[C(t+\la,\la)]=\mbox{const.}\times \prod_{
\stackrel{j,k>0}{1\leq jk\leq N}}
(t-t_{(-j,-k)})^{P(N-jk)},  \eqno(5.2a)
$$ $$
det[C(t+\la,-\la)]=\mbox{const.}\times \prod_{
\stackrel{j,k>0}{1\leq jk\leq N}}(t-t_{(j,k)})^{P(N-jk)}.\eqno(5.2b)
$$
\setcounter{equation}{2}
The first equation tells us that for particular values of $t=t_{(-j,
-k)}$, there are states at the level $jk$ in the Fock space which cannot
be constructed by the Virasoro generators; there are linear combinations
of the states generated by $L^{-J}$ which identically vanish. At the
zeros of (5.2b), we find primary states (``null states" for $c<1$) at
the same level $jk$, which may be constructed by the singular vertex
operators~\cite{KMA}. We note that the vanishing conditions of (5.2a)
and (5.2b) coincide with each other for $c=1 (\la=0)$.

Let us also note a general feature of the BRST formalism.
In sect. 3.3, we discussed the quartet mechanism, where the doublets
appeared in pairs. Suppose a state $|G+1>$ with a ghost number $G+1$
is generated from $|G>$ by the action of the BRST charge $Q_B$:
\EQ
Q_B|G>=|G+1>,\hspace{1cm} Q_B|G+1>=0.
\EN
Then there must be a state $|-G-1>$ which has a nonzero inner product
with $|G+1>$; otherwise $|G+1>$ does not appear in the theory as poles
in Green functions.  A state $|-G>$ defined by the sequence
\EQ
Q_B|-G-1>=|-G>,\hspace{1cm} Q_B|-G>=0
\EN
then has a nonzero inner product with $|G>$ since
\EQ
< -G|c_0|G > = < -G-1|Q_Bc_0|G> = -< -G-1|c_0|G+1>~ =-1.
\EN
Therefore the BRST-doublets always appear in pairs. One can prove that
these quartets do not contribute to the physical spectrum, as in
subsection 3.3.

Combining the properties described in the last paragraphs, we may
understand the origin of discrete states.  The idea is as follows.
Suppose that we have a quartet satisfying (5.3)-(5.5). Let us assume
that in the relation (5.4) the momenta for both matter and gravity
sectors take the values at the zeros of (5.2a). Then one can choose a
state $|-G-1>$ at the level $jk$ such that the vanishing combinations
of Virasoro generators appear in $|-G>$. The state $|-G>$ vanishes
because it is multiplied by a coefficient which has a zero at the
particular values of momenta. We may find a Fock state from $|-G>$ by
dividing out the coefficient (See the examples given below). These two
states no longer belong to a BRST-doublet and do not necessarily
decouple from the physical spectrum; both $|-G-1>$ and $|-G>$ (or
precisely speaking, the corresponding Fock states) are in $Ker~Q_B$. In
order to find them in the physical spectrum, there must be states which
have nonzero inner products with them. From the quartet structure in our
formulation, $|G>$ and $|G+1>$ must be these states. From the
consistency with (5.5), we expect $|G>$ contains the inverse of the
vanishing coefficient. Furthermore we will see that $|G>$ is essentially
a primary state corresponding to a zero of (5.2b). This is to be
expected because any state constructed from the primary states both in
the matter and gravity sectors is in $Ker~Q_B$ since $Q_B$ contains only
Virasoro generators for the matter and gravity. We will explain this
point further in our examples.

Let us study examples at levels one and two.  In order to see how
physical states emerge from the quartet structure, we start from general
momenta and later put them to the values of interest. To do this, we
should first note that the momenta $p^{M,L}=t^{M,L}+\la^{M,L}$ and
$\la^{M,L}$ must satisfy two constraints from the nilpotency of
the BRST charge and the on-shell condition
\bea
&&-\frac{1}{2} [(\la^M)^2 + (\la^L)^2] =1, \nonumber\\
&& {\hat N} + \frac{1}{2} (t^M+\la^M)^2 +
\frac{1}{2}(t^L+\la^L)^2=0.
\ena

At level one ${\hat N}=1$, we have the relations
\bea
Q_B b_{-1}|t+\la>
&=&[L^M_{-1}(\la^M)+ L^L_{-1}(\la^L)]|t+\la>
\nonumber\\
&=&t^M (\a^M_{-1} + {{t^L} \over {t^M}} \a^L_{-1} ) |t+\la>
\ena
\bea
&& Q_B [t^M (1+ ( {{t^L} \over {t^M}})^2)]^{-1}
(\a^M_{-1} + {{t^L} \over {t^M}} \a^L_{-1}) |-(t+\la)> \nonumber\\
&& =-c_{-1} |-(t+\la)>.
\ena
where we have denoted the vacua with momenta $t+\la$ by
\EQ
|t+\la> \equiv |t^M+\la^M>_M \otimes |t^L+\la^L>_L \otimes |0>_{gh}.
\EN
Note that the states in (5.7) and (5.8) have the opposite momenta to
give the nonzero inner product (see the definition in (3.34)). We note
that for general momenta the four states form a quartet. Let us now
choose the momenta $t^M$ close to a zero in (5.2a):
\EQ
t^M=t^M_{(-1,-1)} + \e = \e.
\EN
{}From the relation (5.6) $t^L$ is determined; we choose a solution so
that $t^L=t^L_{(-1,-1)}$ for $t^M=t^M_{(-1,-1)}$.
The ratio $t^L/t^M$ in (5.7) and (5.8) then takes a finite value
\EQ
\frac{t^L}{t^M} = -\frac{\la^M}{\la^L}+O(\e).
\EN
Now it is easy to see that the quartet {\it decomposes} into singlets
when $\e=0$, in a manner we explained earlier. Note that the Fock
states with $N_{FP}=0$ in (5.7) and (5.8) contain only matter
oscillators for $\la^M=0$ ($c=1$), a general feature discussed in
refs.\cite{IT2,BMP}.

We have emphasized the importance of the vanishing combinations of
Virasoro generators on the particular momentum states. Here let us
point out that they are closely related to the ``null states"
corresponding to the zeros in (5.2b). In general, the states which have
nonzero inner products must have opposite momenta. In fact, the state
in (5.8) has opposite momenta $t^M_{(1,1)}$ and $t^L_{(1,1)}$ to those
of the state in (5.7), $t^M_{(-1,-1)}$ and $t^L_{(-1,-1)}$, owing to
the relation $t_{(j,k)}+\la=-t_{(-j,-k)}-\la$. These opposite momenta
precisely correspond to zeros in (5.2b) where we find primary states,
which actually give rise to the state on the LHS of (5.8).

At level two, one can observe the same phenomena. We report only the
result for $\la^M \sim 0$. We take the momenta close to the relevant
zeros at $t=t_{(-1,-2)}$ and $t=t_{(-2,-1)}$:
$$
\e \equiv t^M -t^M_{(-1,-2)}~~ {\rm or} ~~t^M -t^M_{(-2,-1)}.
$$
Expanding the states with respect to $\e$ and $\la^M(\sim 0)$, we find
the quartet structure
\bea
&& Q_B [b_{-2}+b_{-1}(A^ML^M_{-1}+A^LL^L_{-1})]
|t+\la> \nonumber\\
&& =\frac{1+(t_+^M)^2/2}{1+(t_+^M)^2}
\e \{ [\a^M_{-2} \mp {\sqrt 2}(\a^M_{-1})^2] +O(\e,\la^M) \} |t+\la>
\ena
and
\bea
&& Q_B \frac{1+(t_+^M)^2}{6\e(1+(t_+^M)^2/2)}
\{[\a^M_{-2} \pm {\sqrt 2}(\a^M_{-1})^2]
+O(\e,\la^M) \} |-(t+\la)>
 \nonumber\\
&&=[ -\frac{1}{2} (c_{-2} \pm {\sqrt 2} c_{-1} \a^M_{-1})
+O(\e,\la^M) ] |-(t+\la)>
\ena
for the choice of the solution $t^L= \frac{i}{\sqrt{2}}+O(\e,\la^M)$,
where the upper (lower) sign is for $t_{(-1,-2)}$ ($t_{(-2,-1)}$).
The coefficients in (5.12) are given by
\EQ
A^{M,L} = - \frac{1+2 t_+^{M,L} (t^{M,L} - \la^{M,L})}{2
t^{M,L}(t^{M,L}+t^{M,L}_+)} = \mp 1 + O(\e,\la^M),
\EN
where the upper (lower) sign is for matter (gravity) sector. From (5.12)
and (5.13), we again observe the {\it decomposition} of a quartet into
singlets at level two.  On the LHS of (5.13) we find the primary states
corresponding to the zeros $t^M=t_{(1,2)}, t_{(2,1)}$ of
(5.2b).\footnote{Comparison with (2.95) of \cite{KMA} may be useful.}

The coefficients given in (5.14) may be used for general $c$. So one may
find the similar structure for any $c$, starting from the state on the
LHS of (5.12).

It is known that $N_{FP}=0$ discrete states are classified according
to $SU(2)$ generated by $\int dz :e^{\pm i{\sqrt 2}\phi}:$ and $\int
dz i\partial\phi$ ($\phi$ is the scalar field for the matter). From
the ``quartet" structure, we expect that $N_{FP}\ne 0$ discrete states
form $SU(2)$ multiplets as well.  Let us study our examples whether this
is the case for $c^M=1(\la^M=0)$. The state $\a_{-1}^M|(1,1)>$ on the
RHS of (5.7) is a state in the triplet representation, while on the LHS
we find a singlet. In (5.12) and (5.13), we find two states in the
quartet consisting of matter oscillators and the doublet of ghost modes.
These examples suggest the general pattern: ${\bf r}$ in $N_{FP}=0$ goes
into ${\bf r-2}$ in $N_{FP}\ne 0$ under the action of $Q_B$.

The extension of all the above discussions to supersymmetric case
is straightforward, and the origin of the extra physical states may be
explained by the same reasoning.

Returning to the supersymmetric case, we have found extra physical
states with ghost numbers $N_{FP}=0, \pm 1$. We can actually construct
these states for $N_{FP}=0$ as an extension of the bosonic
case~\cite{IT2}:
\bea
(W^+_0)^j|(0,j+k)>_M \otimes |(j+k,0)>_L \otimes |0>_{gh}, \nonumber\\
(W^-_0)^j|(j+k,0)>_M \otimes |(0,-j-k)>_L \otimes |0>_{gh}
\ena
where
\bea
W^{\pm}_0 &=& \int \frac{dz}{2\pi i} W(t_{\pm},z) \nonumber\\
 &=& \int\frac{dz}{2\pi i}:t_{\pm}\psi^M(z) e^{it_{\pm}\phi^M(z)}:
\ena
are the charge screening operators with $t_{\pm}=\pm 1$ for ${\hat c}
= 1$. The matter part of the above states are obtained by the use of
the so-called singular vertex operators~\cite{KMA}.

To see that these states are in the physical spectrum, we first note
that they are in $Ker$ $Q_B$ since $W^{\pm}_0$ commute with $Q_B$
and the rest of the states satisfy the on-shell condition. They
cannot be in $Im$ $Q_B$ since it can be shown that they have
nonzero inner product by using the algebra satisfied by $W^\pm_0$
and a proper definition of the inner product~\cite{IT2}.

Similarly to the bosonic case, the operators in (5.16), together with
$J_0(z)=i\pa\phi(z)$, satisfy an $SU(2)$-like current algebra
\bea
W(t_+,z)W(t_-,w) &\sim& -\frac{1}{(z-w)^2}-\frac{1}{z-w}J_0(w),
\nonumber\\
J_0(z)W(t_{\pm},w) &\sim& \frac{\pm 1}{z-w} W(t_{\pm},w).
\ena
The above discrete states therefore form multiplets with respect
to the global algebra satisfied by the zero modes of the currents.

\vspace{5mm}
In the course of writing this paper, we received ref.~\cite{WIT}
which considerably overlaps with sect. 5 of this paper.

\vspace{1cm}
\noindent{\em Acknowledgement}

We would like to acknowledge the kind hospitality of the Theoretical
Physics Department of Fermilab, where this work was carried out.
Earlier version of the paper contained some argument which was not quite
straightforward for the R sector. We would like to thank the authors
of ref.~\cite{BMP2} for pointing out this.


\newpage
\begin{thebibliography}{99}
\bibitem{BDG} E. Br\'{e}zin and V. Kazakov, \PL{B236} (1990) 144;\\
              M. R. Douglas and S.Shenker, \NP{B335} (1990) 635;\\
              D. J. Gross and A. A. Migdal, \PRL{64} (1990) 127.
\bibitem{BRE} E. Br\'{e}zin, M. Douglas, V. Kazakov and S. Shenker,
              \PL{B237} (1990) 43;\\
              C. Crnkovi\'{c}, P. Ginsparg and G. Moore, \PL{B237}
              (1990) 196.
\bibitem{PGB} G. Parisi, \PL{B238} (1990) 209;\\
              D. J. Gross and N. Miljkovi\'{c}, \PL{B238} (1990) 217;\\
              E. Br\'{e}zin, V. Kazakov and Al. B. Zamolodchikov,
              \NP{B338} (1990) 637;\\
              P. Ginsparg and J. Zinn-Justin, \PL{B240} (1990) 333.
\bibitem{GRO} D. J. Gross and I. Klebanov, \NP{B344} (1990) 475;\\
              D. J. Gross, I. Klebanov and M. Newmann, \NP{B350}
              (1991) 621.
\bibitem{DDK} J. Distler and H. Kawai, \NP{B321} (1989) 509;\\
              F. David, \MPL{A3} (1989) 1651.
\bibitem{SPO} N. Seiberg, \PTPS{102} (1991) 319;\\
              J. Polchinski, Texas preprint, UTTG-19-90(1990); \NP{B324}
              (1989) 123; \NP{B346} (1990) 253; Texas preprint,
              UTTG-06-91(1991).
\bibitem{GTW} A. Gupta, S. Trivedi and M. Wise, \NP{B340} (1990) 475.
\bibitem{BKL} M. Bershadsky and I. Klebanov, \PRL{65} (1990)3088;
              \NP{B360} (1991) 559.
\bibitem{GFK} M. Goulian and M. Li, \PRL{66} (1991) 385;\\
              P. Di Francesco and D. Kutasov, \PL{B261} (1991) 385;\\
              Y. Kitazawa, \PL{B265} (1991) 262;\\
              V. Dotsenko, Paris preprint, PAR-LPTHE 91-18(1991);\\
              N. Sakai and Y. Tanii, TIT preprint, TIT/HEP-168(1991).
\bibitem{GKL} D. J. Gross and I. Klebanov, \NP{B359} (1991) 3.
\bibitem{POL} A. M. Polyakov, \MPL{A6} (1991) 635.
\bibitem{DGR} U. H. Danielsson and D. J. Gross, Princeton preprint,
              PUTP-1258(1991).
\bibitem{IT2} K. Itoh, Texas preprint, CTP-TAMU-42/91(1991).
\bibitem{MMS} S. Mukherji, S. Mukhi and A. Sen, Tata preprint,
              TIFR/TH/91-25 (1991).
\bibitem{IMM} C. Imbimbo, S. Mahapatra and S. Mukhi, Tata preprint,
              TIFR/TH/91-27 (1991).
\bibitem{LIZ} B. H. Lian and G. J. Zuckerman, \PL{B254} (1991) 417.
\bibitem{BMP} P. Bouwknegt, J. M. McCarthy and K. Pilch, CERN preprint,
              CERN-TH.6162/91 (1991).
\bibitem{KOG} M. Kato and K. Ogawa, \NP{B212} (1983) 443.
\bibitem{IKU} K. Itoh, \NP{342} (1990) 449;\\
              T. Kuramoto, \PL{B233} (1989) 363.
\bibitem{MAP} E. Marinari and G. Parisi, \PL{B240} (1990) 375.
\bibitem{DHK} J. Distler, Z. Hlousek and H. Kawai, \IJMP{A5} (1990) 391.
\bibitem{OHT} N. Ohta, \PR{D33} (1986) 1681; \PL{B179} (1986) 347;\\
              J. H. Schwarz, \PTPS{86} (1986) 70.
\bibitem{MIT} M. Ito, T. Morozumi, S. Nojiri and S. Uehara, Prog.
              Theor. Phys. {\bf 75} (1986) 934.
\bibitem{ITO} K. Itoh, \IJMP{A6} (1991) 1233.
\bibitem{TKU} T. Kuramoto, \NP{B346} (1990) 527.
\bibitem{BKT} M. Bershadsky, V. Knizhnik and M. Teitelman, \PL{151B}
              (1985) 31;\\
              D. Friedan, Z. Qiu and S. Shenker, \PL{151B} (1985) 37.
\bibitem{KMA} M. Kato and S. Matsuda, Adv. Studies in Pure Math.
              {\bf 16} (1988) 205.
\bibitem{FF}  B. L. Feigin and D. B. Fuchs, Seminar on Supermanifolds
              No. 5, ed. D. Leites.
\bibitem{FFR} V. S. Dotsenko and V. A. Fateev, \NP{B240} (1984) 312;
              \NP{B251} (1985) 691.
\bibitem{KUO} T. Kugo and I. Ojima, \PTPS{66} (1979) 1.
\bibitem{WIT} E. Witten, IAS preprint, IASSNS-HEP-91-51(1991).
\bibitem{BMP2} P. Bouwknegt, J. M. McCarthy and K. Pilch, CERN preprint,
              CERN-TH.6279/91 (1991).

\end{thebibliography}
\end{document}